\documentclass[oldversion,twocolumn]{aa}
\usepackage{graphicx}
\usepackage{txfonts}
\begin{document}
\title{The nature of damped Lyman $\alpha$ 
and sub-damped Lyman $\alpha$ absorbers}
\author{Pushpa Khare\inst{1}\and
Varsha P. Kulkarni\inst{2}\and
C\'eline P\'eroux\inst{3}\and
Donald G. York\inst{4,5}\and
James T. Lauroesch\inst{6}\and
Joseph D. Meiring\inst{2}}
\institute{Department of Physics, Utkal University,
Bhubaneswar, 751004, India
\and Department of
Physics and Astronomy, University of South Carolina, Columbia, SC 29208,
USA
\and 
European Southern Observatory,
Garching-bei-M\"unchen, Germany
\and 
Department of Astronomy and Astrophysics, University of Chicago, Chicago, IL
60637, USA 
\and Enrico Fermi Institute, University of Chicago, Chicago, IL 60637,
USA
\and
Department of
Physics and Astronomy, University of Louisville, Louisville, KY 40292, 
USA
}

\abstract
{We present arguments based on the measured abundances in individual damped
Lyman $\alpha$ systems (DLAs) and sub-damped Lyman $\alpha$ systems (sub-DLAs),
and also the average abundances inferred in large samples of QSO absorption
line systems, to suggest that the amount of dust in intervening QSO absorbers
is small and is not responsible for missing many QSOs in magnitude
limited QSO surveys. While we can not totally rule out a bimodal dust
distribution with a population of very dusty, metal rich, absorbers which
push the background QSOs below the observational threshold of current optical
spectroscopic studies, based upon the current samples it appears that the
metallicity in QSO absorbers decreases with increase in H I column densities
beyond $10^{19}$ cm$^{-2}$. Thus the sub-DLA population is more metal rich than
the DLAs, a trend which may possibly extend to the non-damped Lyman limit
systems (NDLLS).  Based on the recently discovered mass-metallicity relation
for galaxies, we suggest that most sub-DLAs and possibly NDLLS, are
associated with massive spiral/elliptical galaxies while most DLAs are
associated with low mass galaxies.  The sub-DLA galaxies will then
contribute a larger fraction of total mass (stellar and ISM) and therefore
metals, to the cosmic budget, specially at low redshifts, as compared to the
DLAs.}

\maketitle
\section{Introduction}
Damped Lyman $\alpha$ systems (DLAs), having column density of {H~I} (N$_{\rm
H\;I}$) $\ge\;2\times 10^{20}$ cm$^{-2}$ have been suggested to be the
high-redshift analogs of disks of nearby luminous galaxies. In spite of having
a high amount of neutral hydrogen these absorbers have very little H$_2$
(Ledoux et al. 2003), which raises questions about their being associated with
high star formation activity. A large fraction of DLAs do indeed appear to have
low star formation rates based on deep emission-line imaging searches (e.g.
Kulkarni et al. 2006a and references therein).  Metallicity measurements, using
intermediate to high resolution observations, are available for $>$ 100 DLAs.
These indicate that the metallicity evolution of DLAs is weak; most of the
DLAs, even at $z_{abs}\sim0$ are found to be metal poor (Kulkarni et al. 2005).
It is thus possible that the DLAs may not trace the bulk of star formation in
the Universe and therefore may not be among the leading metal carriers. 

Compilation of observed metallicities of individual DLA systems shows a trend
of decreasing abundance with increasing H I column density (Boisse et al. 1998;
Akerman et al. 2005; Meiring et al. 2006a). The trend possibly exists to lower
H I column densities covering the sub-DLAs, having N$_{\rm H\;I}$ between
10$^{19}$ and $2\times 10^{20}$ cm$^{-2}$ (P\'eroux et al. 2003). A similar
conclusion was drawn by York et al. (2006, hereafter, Y06) based on the average
abundances inferred from the composite spectra of sub-samples drawn from a
sample of 809, intervening Mg II QSO absorption line systems (rest equivalent
width of $\lambda$2796 $>$ 0.3 {\AA} and 1 $<z_{abs}<$ 1.86) compiled from the
Sloan Digital Sky Survey (SDSS) Data Release 1 (DR1). The sub-sample with
highest estimated (near solar) Zn metallicity, was found to have a low average
inferred N$_{\rm H\;I}$ of $\sim10^{20}$ cm$^{-2}$. Several arguments were
given in Y06 to suggest that the DLAs may not be the analogs of the local
bright galaxies, rather, the sub-DLAs and possibly the non-DLA Lyman limit
systems (NDLLS) with N$_{\rm H\;I}$ between 10$^{17}$ and 10$^{19}$ cm$^{-2}$
may be associated with large galaxies. Here we consider this scenario further,
using the currently available data on individual absorbers and the results of
Y06. We consider various selection effects at work and discuss the implications
of the recently established mass-metallicity relation for the nature of DLAs
and sub-DLAs.

In section 2 we study the dependence of the Zn abundance on the column density
of H I, for the observations of individual systems taken from the literature
and for the average values of these quantities obtained by Y06 for large
samples. We present other relevant observational information and discuss the
obscuration threshold and other selection effects to argue for the reality of
the observed dependence.  The implication of this dependence for the nature of
the absorbing galaxies are presented in section 3 and conclusions are presented
in section 4.
\section{Dependence of metallicity on the neutral hydrogen column density}
\subsection{Observational data} The number of metallicity measurements of
individual sub-DLAs has grown considerably since the compilation of P\'eroux et
al. (2003). We have compiled a sample (the literature sample, hereafter TLS) of
all the measurements of Zn abundance of DLAs and sub-DLAs from the literature.
This sample consists of 119 DLAs with $0.1 < z_{abs} < 3.9$ and 30 sub-DLAs
with $0.6 < z_{abs} < 3.2$. The DLA Zn sample is based on our recent HST, MMT,
and VLT data (Khare et al.  2004; Kulkarni et al. 2005; Meiring et al.
2006a,2006b; P\'eroux et al. 2006b) and those from the literature (data
compiled in Kulkarni et al. 2005, and more recent data from Rao et al.  2005;
Akerman et al. 2005; Ledoux et al. 2006). For the sub-DLAs, we compile the data
from Lu et al. (1995, 1996); Pettini et al. (1994, 1999, 2000); Kulkarni et al.
(1999); Ellison \& Lopez (2001); Srianand \& Petitjean (2001);
Dessauges-Zavadsky et al. (2003); Khare et al. (2004); P\'eroux et al. (2006a);
Ledoux et al. (2006); and Meiring et al. (2006a,2006b). The DLA sample contains
68 detections and 51 limits, while the sub-DLA sample contains 13 detections
and 17 limits. Most of these limits are at high redshifts. The full sample used
here is given in Kulkarni et al (2006b).

In Fig. 1, we have plotted [Zn/H], as a function of N$_{\rm H\;I}$ for TLS. The
trend of decreasing abundance with increasing N$_{\rm H\;I}$ is clear.  The
Spearman rank correlation test gives the probability of chance correlation to
be 1.6 $\times 10^{-11}$. It can be seen from the figure that the trend is
similar for systems with $z_{abs}<1.5$ and $z_{abs}>1.5$.

We note that we have only used Zn abundances in Fig.1. There have been studies
which use the Si and S abundances in systems where Zn abundances are not
available due to observational limitations.  We, however, refrain from using
these as (i) given the intrinsically higher abundance and higher strengths of
the detectable absorption lines of these elements, line saturation effects may
be important, leading to an underestimate of their abundances; (ii) Si may be
depleted on the dust grains (up to -1.4 dex in cold ISM); (iii) the lines of Zn
II are detectable to lower redshifts; and (iv) lines of S are often in the
Lyman $\alpha$ forest.  

As noted in the last section, Y06 inferred the average abundances of various
elements in the composite spectra of several sub-samples of SDSS DR1 QSOs with
strong Mg II absorbers. The sub-samples were chosen on the basis of various
properties of the absorption line systems and QSOs. The column densities of
several species, including Zn II, were estimated from inverse
variance-weighted, arithmetic mean, normalized spectra of individual QSOs in
the samples. The geometric mean composite spectra of the same sub-samples were
compared with similar spectra of matching (in emission redshifts and i
magnitudes) samples of QSOs without absorption lines in their spectra, to
determine the absorber rest frame extinction, $E(B-V)$.  The column densities
of neutral hydrogen were estimated by assuming the Small Magellanic cloud (SMC)
dust-to-gas ratio, as the SMC extinction law was found to fit the composite
extinction curves well. In Fig. 1 we have plotted the results for various
sub-samples of Y06. Note that we have used all the sub-samples from Table 1 of
Y06 for which the relevant data were available, some of which were not listed
in their Table A4. Details of their sub-samples which are used here are given
in Table 1. The trend of metallicity dependence on N$_{\rm H\;I}$ is similar to
that in TLS, the Spearman rank correlation test for Y06 points gives the
probability of chance correlation to be 0.098. We have included the main sample
(\#1) of Y06 in our analysis. As the rest of the sub-samples are drawn from
this, its inclusion may cause some bias. We have verified that removing this
sample has very little effect on the correlation noted above.  

Data for individual DLAs show a trend of increasing depletion of Cr with
respect to Zn, and thus, higher dust-to-metal ratio, for higher metallicity
(Ledoux et al. 2003; Akerman et al. 2005; Meiring et al. 2006a). We have
confirmed that a similar trend is shown by the results of Y06. Other
distributions, e.g. [Cr/Zn] vs. log(N$_{\rm Zn\;II}$) and [Cr/Zn] vs.
log(N$_{\rm H\;I}$) are also similar for Y06 and TLS, showing that the results
of Y06, which are average values for large samples, are consistent with the
observations of individual systems.  

We note that the Y06 results are based on the assumption of a constant
dust-to-gas ratio. This ratio may depend on metallicity in view of the
anticorrelation between [Cr/Zn] and [Zn/H] (and in view of Fig.2 below). We
have tried to estimate the effect of this on the relation between [Zn/H] and
N$_{\rm H\;I}$, obtained by Y06, as follows. We assumed that the abundance of
Zn, (X$_{\rm Zn}$) is proportional to N$_{\rm H\;I}^\alpha$. Thus $E(B-V)$
which is proportional to  X$_{\rm Zn}~$N$_{\rm H\;I}$ will be proportional to
N$_{\rm H\;I}^{(1.0+\alpha)}$, giving N$_{\rm H\;I}\propto
E(B-V)^{1.0\over{(1.0+\alpha})}$. Note that the Y06 results assume $\alpha=0$.
For different values of $\alpha$, we determined  N$_{\rm H\;I}$ from the
$E(B-V)$ values, using the above relation. Using this value of H I column
density we determined [Zn/H] and then the slope of the best fit line between
[Zn/H] and Log(N$_{\rm H\;I}$). Note that by construction, this slope should
equal $\alpha$. We found that for positive values of $\alpha$ the slope
remains negative for $\alpha<0.81$ and beyond that remains $<\alpha/2$ until
$\alpha=3$. For negative values of $\alpha$ the slope is somewhat smaller than
the assumed value of $\alpha$ and comes close to it (-0.94) for $\alpha~=-0.9$.
Thus we see that assumption of a metallicity dependent dust-to-gas ratio, in
fact, makes the decrease of Zn abundance with N$_{\rm H\;I}$ even steeper
(slope $\sim$ -0.9) than that (slope = -0.45) found for the assumption of
constant-dust-to-gas ratio (Fig. 1).
\begin{figure}
\includegraphics[width=3.15in,height=3.50in]{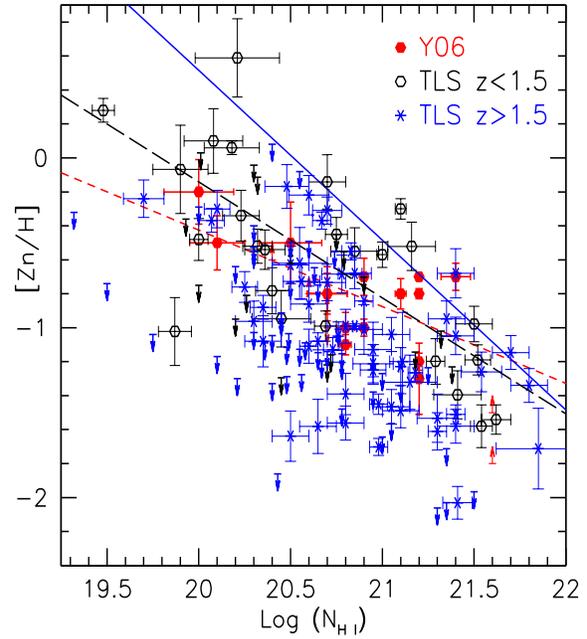}
%\plotone{Fig1.eps}
\caption{Metallicity ([Zn/H]) vs. log(N$_{\rm H\;I}$).  Solid (Red) circles
represent average values for large samples as obtained by Y06. Open (Black)
circles and (blue) stars are for $z_{abs}<1.5$ and $z_{abs}>1.5$, respectively,
for TLS. One sigma error bars are shown.  Also shown are the best fit lines
(long dashed (black) line and short dashed (red) line for TLS and Y06
respectively) obtained by ignoring the limits. The top solid line (blue)
represents the empirical obscuration bias (N$_{\rm Zn\;II} =1.4\times 10^{13}$
cm$^{-2}$).}
\end{figure}
\begin{figure}
\includegraphics[width=3.15in,height=3.50in]{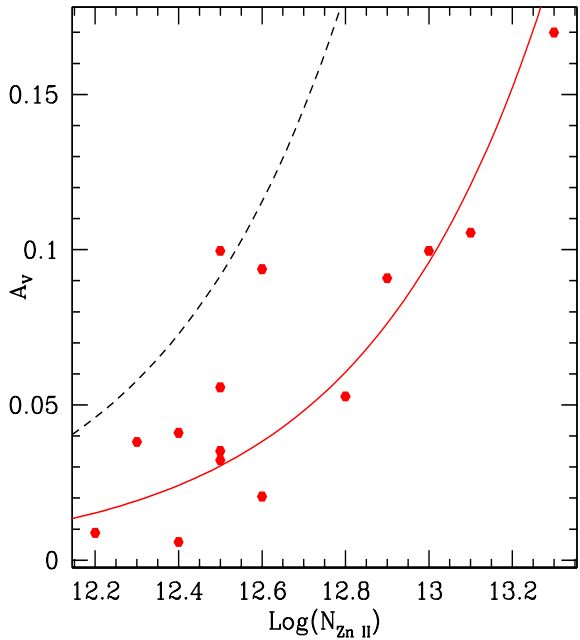}
%\plotone{Fig2.eps}
\caption{Absorber rest frame average extinction, A$_V$, assumed to be
2.93$\times E(B-V)$ vs. log(N$_{\rm Zn\; II}$) for the Y06 results. The linear
best fit relation A$_V$ = 0.1$\times10^{13}$ N$_{\rm Zn\; II}$ is shown as
solid (red) line.  The best fit line obtained for Milky Way sight lines, A$_V$
= 0.3$\times10^{13}$ N$_{\rm Zn\; II}$ by Vladilo and P\'eroux (2005) is shown,
as dashed (black) line for comparison.}
\end{figure}

From an earlier version of this diagram (Fig. 1), with 37 points, including
limits, Boiss\'e et al. (1998) noted the trend of decreasing abundance with
increasing N$_{\rm H\;I}$ (their Fig. 19; hereafter, the Boiss\'e plot).  They
interpreted this as being due to the observational limitations in detecting
weak Zn II lines at the low N$_{\rm H\;I}$ end and an obscuration bias (caused
by dimming of the QSOs, by the dust in the absorbers, below the limit of
magnitude limited surveys) towards the high N$_{\rm H\;I}$ end, causing a
dearth of points in the lower left and upper right corners of the plot
respectively.  Boiss\'e et al. (1998) proposed an obscuration threshold at
N$_{\rm Zn\;II}=1.4\times 10^{13}$ cm$^{-2}$ (above which the background QSO
may be rendered invisible by dust obscuration), shown as a solid (blue)
line in our Fig.  1. This was only an empirical limit based on the sample used
by Boiss\'e et al.  (1998).  It is thus not surprising that a few systems (with
$z_{abs}<1.5$) in Fig. 1 do lie above this threshold. Most of these systems
were observed recently by Khare et al. (2004), Meiring et al.  (2006a,2006b)
and P\'eroux et al. (2006a, 2006b) and many of them have strong Mg II and/or Fe
II lines in the SDSS spectra of the QSO's. Recently, Herbert-Fort et al. (2006)
have reported the presence of two metal strong systems (with $z_{abs}>1.6$) in
SDSS DR3, having Zn II column density greater than the obscuration threshold
(SDSS1610+4724, log(N$_{\rm Zn~II}$)=13.4$\pm{0.03}$ and SDSS1709+3258,
log(N$_{\rm Zn~II}$)=13.19$\pm{0.03}$) and have estimated that $\simeq 5$\% of
the SDSS-DR3 DLA population with $z_{abs}\ge2.2$ in QSOs with $r<19.5$ have
similar Zn II column densities. 
\begin{table*}
\begin{center}
\caption{Relevant data for Y06 sub-samples}
\begin{tabular}{|r|l|l|r|r|r|r|}
\hline
$E(B-V)$&Sample &Selection Criterion$^a$&Number&Log(N$_{\rm Zn~II}$)$^b$&Log(N$_{\rm H~I}$)$^c$\\
(SMC)&number&&of systems&&\\
\hline\hline
  0.002&24&  $\Delta(g-i)^d<$0.2&698&12.4&20.0\\
  0.003&23&  W$_{\rm Mg\;II}^e<$2.0&558&12.2&20.1\\
  0.006&9&  $z_{abs}<$1.3127&404&$>$12.4&20.4\\ 
  0.007&5&  1.53$\le$W$_{\rm Mg\;II}<$1.91&139&12.6&20.5\\
  0.009&11&  i$^f<$19.12&398&$>$12.4&20.6\\
  0.010&13&  $\beta^g<$0.103&405&$<$12.3&20.7\\ 
  0.011&12&  i$\ge$19.12&411&$>$12.4&20.7\\  
  0.011&14&  $\beta\ge$0.103&404&12.5&20.7\\
  0.012&26&  W$_{\rm Mg\;II}\ge$2.5, $\Delta(g-i)<$ 0.2& 97&12.5&20.7\\
  0.013&1&   Full sample & 809&12.3&20.8 \\
  0.014&16&  W$_{\rm Fe\;II}^h$/W$_{\rm Mg\;II}\ge$0.577&369&12.4&20.8\\
  0.018&6&   1.91$\le$W$_{\rm Mg\;II}<$2.52&132 &12.8&20.9\\
  0.019&20&  Fe II $\lambda$2374 present&392&12.5&20.9\\
  0.031&7&   2.52$\le$W$_{\rm Mg\;II}<$5.0&134&12.9&21.1\\
  0.032&8&   W$_{\rm Mg\;II}\ge$2.0&251  & 12.6 &21.2\\
  0.034&17&  W$_{\rm Al\;II}^i$/W$_{\rm Mg\;I}^j<$1.538&85&12.5&21.2\\
  0.034&19&  Fe II $\lambda$2260 present&58  & 13.0&21.2\\
  0.036&21&  Zn II-Mg I $\lambda$2026 present&83&13.1&21.2\\
  0.058&22&  Zn II-Cr II $\lambda$2062 present&31&13.3&21.4\\
  0.081&25&  $\Delta(g-i)\ge$0.2&111 &$>$12.7&21.6\\
  0.085&27&  W$_{\rm Mg\;II}\ge$2.5,$\Delta(g-i)\ge$0.2&48&$>$12.4&21.6\\ 
\hline 				       			 	  
\end{tabular}

\hspace*{1.2in} $^a$ For selection from the full sample (sample 1)\hfill\break
\hspace*{1.2in} $^b$ Estimated from equivalent widths in composite spectra after correction for blends\hfill\break
\hspace*{1.2in} $^c$ Estimated from E(B-V) assuming a constant dust-to-gas ratio\hfill\break
\hspace*{1.2in} $^d$ The difference between the actual colours of QSOs and the
median colours of QSOs at that \hfill\break
\hspace*{1.3in} redshift (Richards et al. 2003)\hfill\break
\hspace*{1.2in} $^e$ Rest equivalent width of Mg II $\lambda$2796 in {\AA}\hfill\break
\hspace*{1.2in} $^f$ i magnitude of the QSO\hfill\break
\hspace*{1.2in} $^g$ Relative velocity w.r.t. the QSO\hfill\break
\hspace*{1.2in} $^h$ Rest equivalent width of Fe II $\lambda$2382 in {\AA}\hfill\break
\hspace*{1.2in} $^i$ Rest equivalent width of Al II $\lambda$1670 in {\AA}\hfill\break
\hspace*{1.2in} $^j$ Rest equivalent width of Mg I $\lambda$2852 in {\AA}\hfill\break

\end{center}			       			 	  
\end{table*}		
\subsection{Establishing the reality of the observed trend}In this section we
present some relevant observational facts and discuss the known selection
effects to suggest that the Boiss\'e plot may indeed not be affected by
these and that the trend of decreasing abundance with increasing
N$_{\rm H\;I}$ may be real.

We first discuss the observed extinction in QSOs and point out that it is much
smaller than that observed in typical Milky Way sight lines and may not cause
significant dimming of the background QSOs.

Vladilo \& P\'eroux (2005) have shown that the extinction A$_{\rm V}$ is
proportional to N$_{\rm Zn\;II}$ for Milky Way sight lines. Their relation is
plotted in Fig. 2. Also plotted is a similar relation obtained for the
sub-samples of Y06.  It can be seen that a correlation does exist between
A$_{\rm V}$ and N$_{\rm Zn\;II}$ for QSO absorbers but that the average
extinction per Zn ion in the these absorbers appears to be smaller than that in
Milky way by a factor of three.  

The maximum value of $E(B-V)$ found by Y06 for their sub-sample (\#27) of most
reddened systems was 0.085 while that for the sub-sample (\#23) of systems with
the rest frame equivalent width of Mg II $\lambda2796 <2.0$ {\AA} is as low as
0.003.  The observer frame A$_{\rm V}$, for the average redshift of 1.33 for
their samples, assuming a 1/$\lambda$ extinction law, is thus smaller than 0.6
and 0.02, respectively for the two samples. These values suggest that the dust
obscuration by the absorbers observed towards the SDSS QSOs, may not be very
important  and may not cause significant decrease in the brightness of the
QSOs.  A similar conclusion is drawn by Murphy \& Liske (2004) from the
analysis of spectral energy distributions of QSOs with DLAs at $z_{abs}\sim3$
in the SDSS DR1. Ellison et al. (2005a) found no significant difference between
the B-K colours of radio selected QSOs with and without DLAs, indicating that
dust obscuration is not very important.  

Below we discuss other indirect evidence indicating that dust obscuration is
not significant in QSOs. 

Wolfe et al. (2005), from the observed depletion pattern of elements, estimated
the fraction of dust obscured QSOs to be $<$ 10\%. Schaye (2001; also see Zwaan
\& Prochaska 2005) has shown that there is an upper limit of $<$ 10$^{22}$
cm$^{-2}$ on the neutral hydrogen column densities of DLAs because of
conversion of {H~I} to H$_2$ and not because of dust obscuration.  Herbert-Fort
et al. (2006) have presented several arguments against the obscuration bias at
redshifts $>1.6 $, most importantly, one based on the magnitude distribution
of the parent QSOs of metal strong systems. 

Samples 11 and 12 of Y06 had i magnitude smaller/larger than 19.1.
Coincidently, this is the cutoff for the SDSS targeting algorithm. Thus the
bright sample consists of QSOs which should be unreddened to enter into the
SDSS QSO catalog while the faint sample consists of candidates which were
observed (and found to be QSOs) on the basis of their being X-ray sources,
radio sources etc. Thus one would expect the faint sample to be more reddened.
However, both samples have similar values of $E(B-V)$ and similar distribution
of $\Delta(g-i)$ values. Thus there is no evidence for higher extinction in
faint SDSS QSOs from the Y06 study. Prochaska et al. (2005) measured
40$\pm$20\% higher gas density in DLAs towards bright QSOs than towards faint
QSOs in SDSS DR3 having DLAs ($z_{abs}>2$). This is contrary to the obscuration
bias and they suggest that gravitational lensing due to DLAs may be important.
Vanden Berk et al. (1997) had found a significant excess of C IV systems in
bright QSOs and interpreted this to be evidence for gravitational lensing. Some
of these C IV systems could however be intrinsic to the QSOs (Richards 2001).
Menard (2005) showed that lensing due to intervening Mg II systems with rest
equivalent width smaller than 1.5, could brighten QSOs by up to -0.2 magnitudes.
Though the two effects (lensing and dust obscuration) have different origins
and need to be understood further, the results of Prochaska et al. (2005)
indicate that the effects of lensing dominate over obscuration effects that may
be present.  

The above arguments seem to indicate that dust obscuration is not
important even in faint QSOs in the SDSS sample. We, however, note that all
these arguments are based on samples of optically selected QSOs and can only
apply if the dust content in QSO absorbers has a continuous distribution. We
note that a few dusty DLAs have been observed (e.g. Junkkarinen et al.  2004;
Motta et al.  2002; Wild et al. 2006) but in none of these cases $E(B-V)$
exceeds 0.42. Such values of $E(B-V)$ may push bright QSOs below the cutoff
(19.1) of SDSS targeting algorithm but these will still be present in the faint
sample of Y06.  Some empty fields have been observed in optical observations
towards radio QSOs by Jorgenson et al. (2006) which could potentially be
obscured by dusty absorbers, though the authors have argued against such a
possibility on the basis of the DLA statistics observed in radio selected QSOs. 

Some theoretical studies (e.g. Vladilo and P\'eroux 2005) have argued that up
to 30\% to 50\% of the QSOs may be missed in magnitude limited surveys due to
dust obscuration.  Smooth-particle-hydrodynamics simulations do need to
introduce dust obscuration to explain the observed (low) DLA metallicity (Cen
et al. 2003). Our arguments above do not rule out the possibility of a bimodal
distribution of dust columns, with a population of dusty absorbers which have
pushed the background QSOs below the observational limit of the optical
surveys.  Though this possibility can only be verified with observations of
fainter QSOs, no compelling evidence for such a population is found from the
study of metallicity, column density distribution and mass density of H I in
DLAs in radio selected samples of QSOs (Akerman et al. 2005; Jorgenson et al.
2006). Their study suggests that the optically selected samples give a fair
census of the population of DLA absorbers.  It should, however, be noted that
these samples are still small and cover redshifts $>$ 1.86 only. 

From all the above arguments, it seems very likely that not
many points in the upper right corner of the Boiss\'e plot are missed due to
dust obscuration.  

The missing points in the lower left corner of the Boiss\'e plot have been
attributed to an artifact of the sensitivity of observations (Boiss\'e et al.
1998). A few systems may indeed be present in this region as seen from the few
upper limits there in Fig. 1. However, it may be noted that in the analysis of
Y06, solar metallicity was estimated in the sub-sample (\#24) of 698 systems
with average N$_{\rm H\;I}$ of about 10$^{20}$ cm$^{-2}$. Most (638) of these
systems did not have detectable Zn II lines in their individual SDSS spectra.
We note that even a solar metallicity system with N$_{\rm H\;I}$ of about
10$^{20}$ cm$^{-2}$, may not produce detectable Zn II lines in an individual
SDSS spectrum which typically has S/N $\sim$ 10-15 and a 3$\sigma$ detection
limit in the observer frame of $>$ 0.3 {\AA}. So the absence of Zn II lines in
an individual SDSS spectrum does not mean low Zn abundance. The composite
spectrum of 698 systems has S/N higher than that in a typical SDSS spectrum by
a factor of $\sim$ 25-30, enabling the detection of Zn II lines. In the
sub-samples of Y06 (\#s 21 and 22) comprised of systems with detectable Zn II
and Cr II lines, the metallicity was estimated to be lower than that in the
sample of 698 systems.  It thus appears that their metallicity measurements,
being averages over large samples, are not affected by observational
limitations in detecting weak lines. We also note that the average solar
metallicity in the sub-sample \# 24 can not be an artifact of a possible high
metallicity tail of the metallicity distribution in sub-DLAs.  The composite
spectrum, used for abundance determination is the arithmetic mean of individual
spectra. Thus, if 90\% of the 698 systems had subsolar abundance with [Zn/H]=-1
and only 10\% had solar abundance then, assuming an H I column density of
10$^{20}$ cm$^{-2}$ for the systems, the average equivalent width of
$\lambda2026$ line of Zn II will be $0.9\times0.007+0.1\times0.07=0.013
\AA$, which will correspond to highly sub-solar abundance of Zn II for
this sub-sample as opposed to the solar abundance derived by Y06. It
may also be noted that the sub-samples \#23 and 24, do have a large
number of systems with low N$_{\rm H~I}$ which in principle, could have
had low [Zn/H] and could then have been present in the lower left corner
of the Boisse plot. Sub-sample \#23 which is the sample of systems with
W$_{\rm Mg~II} < 2.0$ {\AA}, includes all the systems in sub-samples \#s
2, 3 and 4 (with W$_{\rm Mg~II}<1.91$ {\AA}, all together 397 systems)
which have $E(B-V)<0.001$ and thus N$_{\rm H~I} < 5\times 10^{19}$
cm$^{-2}$. Similarly sub-ample \#24 has 698 systems. Even assuming it has
all the systems with W$_{\rm Mg~II}>1.91$ {\AA} (i.e.  all the systems
not included in sub-amples \#s 2, 3 and 4) from the original sample of
809 systems, it will still have $\ge 286$ systems with N$_{\rm H~I}
<5\times 10^{19}$ cm$^{-2}$. However, the high values of the average
abundances for these sub-samples suggest that most of these low N$_{\rm
H~I}$ systems do not have low abundance and thus do not occupy the lower
left corner of the Boisse plot. We also note that it is possible that
these two sub-samples do include some DLAs. However, as argued above, the
DLAs do not have high abundances and can only reduce the average
abundances of the sub-samples, thereby, making the case for higher
abundance of sub-DLAs even stronger. 

On the conservative side, we point out that even if some points (in TLS) are
indeed missing in the lower left region of Fig. 1 due to sensitivity limit of
observations, the average metallicity at low N$_{\rm~H~I}$ (in TLS) will still
be higher than that at high N$_{\rm~H~I}$ due to the large number of high
metallicity, low N$_{\rm~H~I}$ systems that have been observed (assuming the
obscuration bias to be absent, as argued above). It thus appears that the
sub-DLA metallicity is indeed higher than that of DLAs and is close to the
solar value, at low redshifts. 

It thus seems very likely, that the observed trend of
decreasing abundance with increasing N$_{\rm H\;I}$ is not
due to selection effects and is real and that DLAs 
are not among the major metal carriers in the universe. We note that
super-solar abundances have been observed in seven sub-DLAs (Pettini et al.
2000; Khare et al. 2004; Prochaska et al. 2006; Meiring et al. 2006b).  Near
solar abundance has been estimated in one NDLLS (Jenkins et al.  2005).
Bergeron et al.  (1994), estimated abundances ([X/H]), of several Mg II Lyman
Limit systems with low Lyman limit optical depth, with redshifts between 0.1
and 1.1, to be between -0.5 and -0.3 dex.  Super-solar abundances have also
been estimated for three systems with redshifts between 0.7 and 0.95 (Charlton
et al.  2003, Ding et al.  2003; Meseiro et al. 2005) and near-solar abundance
has been estimated in one system at redshift of 0.064 (Aracil et al.  2006);
all these systems have N$_{\rm H\;I}<10^{16}$ cm$^{-2}$. We however caution
that the derived abundance in NDLLS and systems with smaller H I column
densities do depend on the details of photoionization calculations.  

Having argued above, for the reality of the observed dependence of Zn
metallicity on H I column density, below, we consider its implications. In
particular, we consider consequences of the hypothesis that the sub-DLAs, and
possibly NDLLS, have higher abundances than DLAs and represent a selection of
galaxies that are the major metal carriers in the universe.  
\section{Implications}
Recently a mass-metallicity relation has been discovered by several groups.
Tremonti et al. (2004), from the imaging and spectroscopy of ~ 53,000
star-forming galaxies at $z \sim$ 0.1, found a tight ($\pm 0.1$ dex)
correlation between stellar mass and metallicity spanning over 3 orders of
magnitude in stellar mass and a factor of 10 in metallicity. Savaglio et al
(2005), from a sample of 56 galaxies identified a strong correlation between
mass and metallicity at 0.4 $< z <$ 1.0.  They predict that the generally metal
poor DLA galaxies have stellar masses of the order of 10$^{8.8}$ M$_\odot$
(with a dispersion of 0.7 dex) from $z$=0.2 to $z$=4. Erb et al.  (2006) have
obtained a mass-metallicity relation in a sample of 87 galaxies at $<z>\;\sim$
2.26 which is similar to the relation for local galaxies (Tremonti et al. 2004)
except for an offset of 0.3 dex in metallicities, indicating that galaxies of a
given mass have lower metallicity at high redshift. They however, note that the
uncertainty in the metallicity offset between the $z\sim2$ and local galaxies
is approximately a factor of 2, about the same as the offset itself. Ledoux et
al. (2006) found a correlation between the metallicity and velocity widths of
lines of low ionization species over two orders of magnitude in metallicity, at
1.7 $< z <$ 4.3. Assuming velocity widths to be a measure of mass, their
mass-metallicity relation is consistent with that found for local galaxies by
Tremonti et al. (2004). We, however, note that Bouche et al. (2006) found an
anticorrelation between halo mass and Mg II $\lambda$ 2796 equivalent widths
which seems to go against the results of Ledoux et al. (2006).  

At low redshifts the metallicity of most DLAs is almost an order of magnitude
lower than the solar value, while, as suggested by results of Y06 and those of
Kulkarni et al. (2006b), the mean sub-DLA metallicity seems to be close
to the solar value. {\it The mass-metallicity relation of Tremonti et al.
(2004) would imply stellar masses of about 10$^{11}$ M$_\odot$ and $<10^{9}$
M$_\odot$ respectively for the sub-DLAs and DLAs.} These numbers will be
smaller by a factor of 2 if the mass-metallicity relation of Erb et al.  (2006)
is used. If the stellar metallicity is lower/higher than the gas phase
metallicity (as determined by the abundances in DLAs and sub-DLAs) the masses
of the DLA and sub-DLA galaxies may be correspondingly lower/higher, however,
the ratio of the two masses will not be affected.

We thus propose that the sub-DLAs and DLAs in general represent different types
of galaxies. By and large, the sub-DLAs are produced by massive galaxies, with
higher metallicity, while the DLAs are produced by less massive galaxies with
lower metallicity. A few sub-DLAs may indeed arise from lines of sight which
encounter low H I columns through DLA galaxies and therefore, have lower
metallicity. These will give rise to points in the lower left region of Fig. 1.
Similarly, a few DLAs may arise in lines of sight through sub-DLA galaxies and
may have higher metallicities though the probability of this happening may be
small. In our Galaxy, for instance, clouds with log(N$_{\rm H~I}$)$>$20.3 are
very small and represent only a tiny part of the cross section of the entire
Galaxy.

It has been suggested that the low metallicity found in DLAs is caused by
metallicity gradients. Differences between emission line and absorption line
metallicities have been observed in a few cases (Chen et al. 2005; Ellison et
al. 2005b) but not in others (Schulte-Ladbeck et al.  2005; Bowen et al.
2005).  The metallicity gradients in nearby spirals are fairly weak (Bresolin
et al.  2004) and can not explain the low metallicity in DLAs with small impact
parameters (e.g. Kulkarni et al. 2005). It has been suggested in several
studies (e.g.  Kauffmann 1996; Das et al. 2001; Zwaan et al. 2005) that DLAs
may result for small values of impact parameters while larger impact parameters
through the same absorbers may give rise to NDLLS and Lyman $\alpha$ forest
systems. In this case, abundance gradients can not be invoked to explain lower
abundances in DLAs as compared to sub-DLAs/NDLLS. This scenario has also not
been verified observationally.  

Kulkarni et al. (2006b) have shown that the N$_{\rm H~I}$ weighted mean
metallicity for sub-DLAs is a factor of 6 higher than that for DLAs at low
redshifts.  They estimate that at these redshifts, the contribution of the ISM
in sub-DLAs to the cosmic metal budget may be several times that of ISM in
DLAs. Prochaska et al. (2006), based on their observations of two metal strong
sub-DLAs (Super LLS in their paper) and results from their sub-DLA survey,
estimated that the ISM in sub-DLAs may contribute at least 15\% to the metal
budget of the universe at $z \simeq2$. 

{\it If the galaxies that give rise to sub-DLAs are indeed more massive than the
galaxies that give rise to DLAs, as suggested above, then an individual sub-DLA
galaxy will contribute more to the stellar mass of the universe than an
individual DLA galaxy.}  Also, in CDM cosmology, as the massive galaxies form by
mergers of smaller galaxies which triggers star formation leading to a higher
rate of metal production, they are not only expected to show a higher rate of
metallicity evolution (as is found by Kulkarni et al. 2006b) but are also
expected to contribute more mass at lower redshifts. {\it It is thus possible
that the contribution of sub-DLA producing galaxies to the cosmic metal budget
at lower redshifts may indeed be considerably higher than 15\% (as estimated by
Prochaska et al. (2005) at $z \simeq2$) and may help alleviate the missing
metals problem.}

Boissier et al. (2003) compared the observed properties of DLAs with the
predictions of simple models of present day disk galaxies and showed that low
surface brightness galaxies contribute as much as spirals to the number and H I
mass of DLAs. Zwaan et al. (2005) have shown that properties of DLAs are
consistent with their forming in galaxies of various morphological types, with
87\% of DLA cross-section being contributed by sub-L$_*$ galaxies.
Semi-analytic models indicate sub-L$_*$ galaxies to be major contributors to
DLA cross-section (Okoshi \& Nagashima 2005). Thus there seems to be some
evidence for a significant fraction of DLAs to be associated with sub-L$_*$
galaxies. 

Deep imaging has shown low redshift DLAs to be often associated with dwarf
galaxies (Rao et al. 2003). Le Brun et al. (1997) suggested that the DLA
galaxies (which were selected on the basis of a damped Lyman $\alpha$ line or 21
cm absorption or a very high Mg II/Fe II ratio) strongly differ from the Mg II
selected galaxies, the latter being mostly luminous galaxies with evidence of
recent star formation activity.  Indeed emission-line imaging searches suggest
that a large fraction of DLAs appear to have low star formation rates (Kulkarni
et al. 2006a and references therein). Deep imaging of three sub-DLA galaxies at
$z\;<$ 0.7 shows them to be associated with L$>0.6$ L$_*$, disk/spiral galaxies
(Zwaan et al.  2005).  Note that the lines of sight through our Galaxy are,
statistically speaking, mostly metal rich sub-DLAs with N$_{\rm H\;I}
<2\times10^{20}$ cm$^{-2}$, perpendicular to the plane. All these support our
hypothesis about the nature of DLA and sub-DLA galaxies. Chen and
Lanzetta (2003), however, found the DLA galaxies to have a variety of
morphologies. Djorgovski et al. (1996) and Moller et al. (2002, 2004) have
observed a few high-redshift ($z\ge$2) absorbing galaxies in Lyman $\alpha$
emission and found their properties to be similar to disk/Lyman break galaxies.
Chun et al. (2006), using adaptive optics imaging for 4 DLAs and 2 sub-DLAs at
$z<$ 0.5, found most of the candidate absorber galaxies or their companions to
have low-luminosity ( $<$ 0.1 L$_*$).

Thus, it seems that the results of some of the DLA and sub-DLA imaging
studies are consistent with the hypothesis we make here. More systematic
imaging surveys are needed to confirm the ideas presented in this paper. Deep
imaging in K-band for absorbers with high abundances at $z_{abs}<1.5$ should be
able to confirm the presence of massive (red) galaxies, while imaging searches
in narrow, optical emission lines should help in confirming that DLAs are
mostly dwarf galaxies.

\section{Conclusions}
We have studied the dependence of metallicity on N$_{\rm H\;I}$ in QSO
absorbers with N$_{\rm H\;I}>10^{19}$ cm$^{-2}$ and have discussed various
selection effects that may be giving rise to the observed trend. We
have argued that the selection effects are not important and that the observed
trend is real. Our conclusions based on these arguments, and subject to
confirmation by future observations, are as follows: 
\begin{enumerate}
\item{} The amount of dust in QSO absorbers is small and is not
responsible for missing many QSOs in magnitude limited surveys. We can not,
however, rule out the possibility of a bimodal distribution of dust columns
such that there may exist a population of dusty absorbers which push the
background QSOs below the observational threshold of current optical
spectroscopic studies and is completely invisible.
\item{} The metallicity in QSO absorbers with N$_{\rm H\;I}>10^{19}$ cm$^{-2}$,
decreases with increase in H I column density of these absorbers. The trend
possibly continues to lower H I column densities.  The sub-DLAs thus have
higher metal abundances as compared to the DLAs at redshifts between 0 and 2.
\item{} The observed mass-metallicity relation suggests that most DLAs
are associated with low mass ($<10^9$ M$_\odot$) galaxies while most
sub-DLAs are associated with massive spiral/elliptical galaxies.  It is
possible that the non-DLA LLS may also be metal rich and may be associated with
massive galaxies. 
\item{} The sub-DLA galaxies will contribute a larger fraction of total
mass (stellar and ISM) and therefore metals, to the cosmic budget, specially at
low redshifts, as compared to the DLAs. The Sub-DLAs and possibly, non-DLA LLS
galaxies, may contain a much larger fraction of the metals at $z< 1$ than has
been appreciated.
\item{} The few imaging studies, of galaxies responsible for quasar absorption
line systems done so far, are ambiguous on the morphology of DLA and sub-DLA
galaxies. More systematic, deep imaging in r-band, Ks-band and in narrow
emission lines is essential to confirming the inferences of this paper.
\end{enumerate}
\begin{acknowledgements}
PK acknowledges support from the Department of Science and Technology,
Government of India (SP/S2/HEP-07/03). VPK and JDM acknowledge support from the
U.S. National Science Foundation grant AST-0206197.
\end{acknowledgements}


\begin{thebibliography}{}
\bibitem{Ak05} Akerman, C. J., Ellison, S. L., Pettini, M., \& Steidel, C. C. 2005,
A\&A, 440, 449
\bibitem{Ar06} Aracil, B., Tripp, T. M., Bowen, D. V., Proschaska, J. X., Chen,
H. W., \& Frye, B. L. 2006, MNRAS, 367, 139
\bibitem{Be94} Bergeron, J. et al. 1994, ApJ, 436, 33
\bibitem{Boi98} Boiss\'e, P., Le Brun, V., Bergeron, J., \& Deharveng, J. M. 1998,
A\&A, 333, 841
\bibitem{BPM01} Boissier, S., P\'eroux, C. \& Pettini, M. 2003, MNRAS, 338, 131
\bibitem{Bow05} Bowen, D. V., Jenkins, E. B., Pettini, M., \& Tripp, T. M.
2005, ApJ, 635, 880
\bibitem{Bou06} Bouch\'e, N., Murphy, M. T., P\'eroux, C., Csabai, I., \& Wild,
V. 2006, MNRAS, 371, 495
\bibitem{Bre04} Bresolin, F., Garnett, D. R., \& Kennicutt, R. C. 2004, ApJ, 615, 228
\bibitem{Cen03} Cen, R., Ostriker, J. P., Prochaska, J. X., \& Wolfe, A. M.
2003, ApJ, 598, 741
\bibitem{Cha03} Charlton, J. C., Ding, J., Zonak, S. G., Churchill, C. W.,
Bond, N. A., \& Rigby, J. R. 2003, ApJ, 589, 111
\bibitem{Ch03} Chen, H. W., \& Lanzetta, K. M. 2003, ApJ, 597, 706
\bibitem{Ch05} Chen, H. W., Kennicutt, R. C., \& Rauch, M. 2005, ApJ, 620, 703
\bibitem{Chu06} Chun, M. R., Gharanfoli, S., Kulkarni, V. P., \& Takamiya, M.
2006, AJ, 131, 686
\bibitem{Da01} Das, S., Khare, P., \& Samantray, A. 2001, A\&A, 373, 843
\bibitem{Dess03} Dessauges-Zavadsky, M., P\'eroux, C., Kim, T.-S., D'Odorico,
S., \& McMahon, R. G. 2003, MNRAS, 345, 447
\bibitem{Di03} Ding, J., Charlton, J. C., Churchill, C. W.,\&  Palma, C. 2003,
ApJ, 590, 746
\bibitem{Dj96} Djorgovski, S. G., Pahre, M. A., Bechtold, J., \& Elston, R.
1996, Nature, 382, 234
\bibitem{EL01} Ellison, S. L., Lopez, S. 2001, A\&A, 380, 117
\bibitem{EL05a} Ellison, S. L., Hall, P. B. \& Lira, P. 2005a, ApJ,  
\bibitem{EL05b} Ellison, S. L., Kewley, L. J., Mallen-Orleans, G. 2005b, MNRAS,
357, 354
\bibitem{ER06} Erb, D. K., Shapley, A. E., Pettini, M., Steidel, C. C., Reddy,
N. A., \& Adelberger, K. L. 2006, ApJ, 644, 813 
\bibitem{Her06} Herbert-Fort, S., Prochaska, J. X., Dessauges-Zavadsky, M.,
Ellison, S. L., Howk, C., Wolfe, A. M., \& Prochter, G. E. 2006, PASP, 118,
1077
\bibitem{Jen05} Jenkins, E. B., Bowen, D. V., Tripp, T. M., \& Sembach, K. R.
2005, ApJ, 623, 767
\bibitem{Jor06} Jorgenson, R. A., Wolfe, A. M., Prochaska, J. X., Lu, L., Howk,
J. C., Cooke, J., Gawaiser, E., \& Gelino, D. M. 2006, ApJ, 646, 730
\bibitem{Ju04} Junkkarinen, V. T., Cohen, R. D., Beaver, E. A., Burbidge,
E. M., Lyons, R. W. \& Madejski, G. 2004, {ApJ}, 614, 658 
\bibitem{Ka96} Kauffmann, G. 1996, MNRAS, 281, 475 
\bibitem{Kh04} Khare P., Kulkarni V. P., Lauroesch J. T., York D. G, Crotts P.
S., \& Nakamura O. 2004, ApJ, 616, 86 
\bibitem{Kul99} Kulkarni, V. P., Bechtold, J.,
 Ge, J. 1999, in Proc. of ESO Conference on ``Chemical Evolution from Zero to
High Redshifts'', eds. M. Rosa and J. Walsh, (Springer-Verlag), 275
\bibitem{Kul05} Kulkarni, V. P., Fall, S. M., Lauroesch, J. T., York, D. G,
Welty, D. E, Khare, P., \& Truran, J. W. 2005, ApJ, 618, 68
\bibitem{Kul06a} Kulkarni, V. P., Woodgate, B. E., York, D. G., Thatte, D. G.,
Meiring, J. D., Palunas, P., \& Wassell, E. 2006a, ApJ, 636, 30
\bibitem{Kul06a} Kulkarni, V. P., Khare, P., P\'eroux, C., York, D. G.,
Lauroesch, J. T., \& Meiring, J. D. 2006b, submitted to ApJL, astro-ph/0608126
\bibitem{Lb97} Le Brun, V., Bergeron, J., Boiss\'e, P., \& Deharveng, J. M.
1997, A\&A, 321, 733 
\bibitem{LPS03} Ledoux, C., Petitjean, P., \& Srianand, R. 2003, MNRAS, 346,
209
\bibitem{Le06} Ledoux, C., Petitjean, Fynbo, J. P. U., Moller, P., \& Srianand,
R.  2006, A\&A, 457, 71L
\bibitem{LU95} Lu, L., Savage, B. D., Tripp, T. M., Meyer, D. M. 1995, 
 ApJ, 447, 597
\bibitem{LU96} Lu, L., Sargent, W. L. W., Barlow, T. A., Churchill, C. W., 
Vogt, S. S. 1996, ApJS, 107, 475
\bibitem{Mas05} Maseiro, J. D. R., Charlton, J. C., Ding, J., Churchill, C. W.,
\& Kacprzak, G. 2005, ApJ, 623, 57
\bibitem{Mei06a} Meiring, J. D., Kulkarni, V. P., Khare, P., Bechtold, J., York,
D.  G., Cui, J., Lauroesch, J. T., Crotts, A. P. S., \& Nakamura, O. 2006a,
MNRAS, 370, 43
\bibitem{Mei06b} Meiring, J. D., Lauroesch, J. T., Kulkarni, V. P., P\'eroux,
C., Khare, P., York, D.  G. 2006b, MNRAS, submitted
\bibitem{Men05} Menard, B. 2005, ApJ, 630, 28
\bibitem{Mol02} Moller, P., Warren, S. J., Fall, S. M., Fynbo, J. U., \&
Jakobsen, P. 2002, ApJ, 574, 51 
\bibitem{Mol04} Moller, P., Fynbo, J. P. U., \& Fall, S. M. 2004, A\&A, 422, L33
\bibitem{Mo02} Motta,  V., Mediavilla, E., Muñoz, J. A., Falco, E., Kochanek,
C. S., Arribas, S., García-Lorenzo, B., Oscoz, A., \& Serra-Ricart, M. 2002,
ApJ, 574, 719
\bibitem{Mur04} Murphy, M. \& Liske, J. 2004, MNRAS, 354, 31
\bibitem{Ok05} Okoshi, K., \& Nagashima, M. 2005, ApJ, 623, 99 
\bibitem{Per03} P\'eroux, C., Dessauges-Zavadsky, M., D'Odorico, S., Kim,
T. S., \& McMahon, R. G. 2003, MNRAS, 345, 480
\bibitem{Per06a} P\'eroux, C., Kulkarni, V. P., Meiring, J. D., Ferlet, R.,
Khare, P., Lauroesch, J. T., Vladilo, G., \& York, D. G. 2006a, A\&A, 450, 53
\bibitem{Per06b} P\'eroux, C., Meiring, J. D., Kulkarni, V.P., Ferlet, R.,
Khare, P., Lauroesch, J. T., Vladilo, G., \& York, D. G. 2006b, MNRAS, 372, 369
\bibitem{Pet94} Pettini, M., Smith, L. J., Hunstead, R. W., King, D. L.
1994, ApJ, 426, 79
\bibitem{Pet99} Pettini, M., Ellison, S. L., Steidel, C. C., Bowen, D. V.
1999, ApJ, 510, 576
\bibitem{Pet00} Pettini, M., Ellison, S. L., Steidel, C. C., Shapley, A. E., \&
Bowen, D. V. 2000, ApJ, 532, 65
\bibitem{Pro05} Prochaska, J. X., Herbert-Fort, S., \& Wolfe, A. M. 2005, ApJ,
635, 123
\bibitem{Pro06} Prochaska, J. X., O'Meara, J. M., Herbert-Fort, S., Burles, S.,
Prochter, G. E., \& Bernstein, R. A. 2006, astro-ph/0606573
\bibitem{Rao03} Rao, S. M., Nestor, D. B., Turnshek, D. A., Lane, W. M.,
Monier, E. M. \& Bergeron, J. 2003, ApJ, 595, 94  
\bibitem{Rao05} Rao, S. M., Prochaska, J. X., Howk, C., \& Wolfe, A. M. 2005,
AJ, 129, 9
\bibitem{Ri01} Richards, G. T. 2001, ApJS, 133, 53
\bibitem{Ric03} Richards, G. T. et al. 2003, AJ, 126, 1131
\bibitem{Sav05} Savaglio, S.  et al. 2005, ApJ, 635, 260
\bibitem{Sc01} Schaye, J. 2001, ApJ, 562, L95
\bibitem{Sch05} Schulte-Ladbeck. R. E., Konig, B., Miller, C. J., Hopkins, A.
M., Drozdovsky, I. O., Turnshek, D. A., Hopp, U. 2005, ApJ, 625, L79
\bibitem{Sr01} Srianand, R., Petitjean, P. 2001, A\&A, 373, 816
\bibitem{Tre04} Tremonti, C. A. et al. 2004, ApJ, 613, 898
\bibitem{Vb97} Vanden Berk, D. E., Quashnock, J. M., York, D. G., Yanny, B.
1997, ApJ, 469, 78
\bibitem{Vla04} Vladilo, G., \& P\'eroux, C. 2005, A\&A, 444, 461
\bibitem{Wi06} Wild, V., Hewett, P. C., \& Pettini, M. 2006, MNRAS, 367, 211
\bibitem{Wol05} Wolfe, A. M., Gawiser, E., \& Prochaska, J. X. 2005, ARAA, 43,
861
\bibitem{Yo06} York, D. G., et al. 2006, MNRAS, 367, 945
\bibitem{Zwa06} Zwaan, M. A., \& Prochaska, J. X. 2006, ApJ, 643, 675 
\bibitem{Zwa05} Zwaan, M. A., van der Hulst, J. M., Briggs, F. H., Verheijen,
M. A. W., \& Ryan-Weber, E. V. 2005, MNRAS, 364, 1467 
\end{thebibliography}
\end{document}